\documentclass[aps,prl,twocolumn,superscriptaddress,nofootinbib,nobalancelastpage,showpacs,nobibnotes]{revtex4}
\usepackage{epsfig}

\begin{document}
\title
{Hydrodynamics of an electron-positron plasma near a black hole; applications to
jet formation.   
}

\author{R. F. Sawyer}
\affiliation{Department of Physics, University of California at
Santa Barbara, Santa Barbara, California 93106}

\begin{abstract}

We investigate some features of the  hydrodynamics and neutrino physics in the (predominantly)
electron-positron plasma above a hyperaccreting disk (or torus) around a black hole, a conjectured
engine for a short gamma ray burst. We suggest a possible scenario in which plasma in the region
very near the black hole, energetically driven by neutrino annihilation, emerges as a subsonic wind,
which in a spherically symmetrical case would decelerate as it moves out. In this case we argue that
the plasma heating will be primarily through neutrino-electron and neutrino-positron scattering,
and that this process will be important throughout a  region considerably larger than that of the 
neutrino annihilation process. In simple solutions a relatively  gentle anisotropy in the heating 
through this process can create an approximately conical sonic surface, aligned with the system's axis. Inside this cone the
fluid accelerates upwards as in standard jet models.


\end{abstract}
\maketitle

\section{Introduction}
A popular model for the mechanism of short gamma ray bursts
begins with a ``hyperaccretion disk" (or torus) created in the last moments of a 
neutron star-black hole merger \cite{popham}-\cite{janiuk}. An electron pair plasma is then created in a region of relatively low density along the disk axis through the process $\nu+\bar \nu \rightarrow e^+ +e^-$,
where the neutrinos originate in the dense inner part of the torus. This pair production
is concentrated in a region very near the black hole, since it is this region that 
combines proximity to the neutrino production region with the geometric advantage
that is given when neutrino beams originating from opposite sides of the torus
meet head-on in the region of the axis. Indeed, in the simulations reported in ref.
\cite{aloy} the energy density deposited in the plasma per unit time near the
axis is proportional to $z^{-5}$ where $z$ is the distance to the center of the black
hole. 

The plasma is
then to be accelerated, by hydrodynamics alone, to form a $\Gamma \sim 100$
jet as it moves out the disk axis.

In the case of high rates of accretion, the densities and temperatures of the matter
in the dense torus render the matter opaque, or somewhat opaque, to neutrinos in the innermost region. Thus
both the disk dynamics and the proposed $e^\pm$ production mechanism depend
on a neutrino transport calculation. Depending on choices of input parameters, as well
as on the computational framework for estimating production rates, one may arrive at
the conclusion that the model is sufficient to supply the jet for an energetic gamma ray burst,
or at the conclusion that it is insufficient.

In the present note we propose an alternative to this
picture, while adhering to the general description that an $e^\pm$ plasma formed on the periphery
of the torus can be accelerated by hydrodynamic forces to form an extremely relativistic jet
along the direction of the disk axis. Taking a three solar mass black hole, as in the
simulations of ref. \cite{aloy}, with a Schwarzchild radius of approximately 7 km.,
let us circumscribe the system with a sphere of radius, say, 50 km. and ask 
what happens in this region. In the models used in ref.\cite{aloy}, the background density
at this distance is fairly small, even at polar angles $\theta$ up to $60$ deg.; in model
B of this reference it ranges from about $10^6 g\,\, cm^{-3}$ along to axis to about 
$10^8 g\,\, cm^{-3}$ for $\theta$ =60 deg.

We shall describe a scenario in which this whole region
fills with $e^\pm$ plasma, with small outflow velocity, in which the variation of
plasma temperature with polar angle drives a subsonic$\rightarrow$ supersonic transition 
along a more or less conical surface, as shown in fig.1.
Looking at fig.1 we should state at the outset that some of the features are calculated
later in this paper, and some imagined. Two key elements that are calculated or estimated are the 
exhibited ratios of temperatures and the shape of the sonic transition
surface.

\begin {figure}[ht]
    \begin{center}
       \epsfxsize 2.75in
        \begin{tabular}{rc}
           \vbox{\hbox{
$\displaystyle{ \, { } }$
               \hskip -0.1in \null} %\vskip 0.2in
} &
            \epsfbox{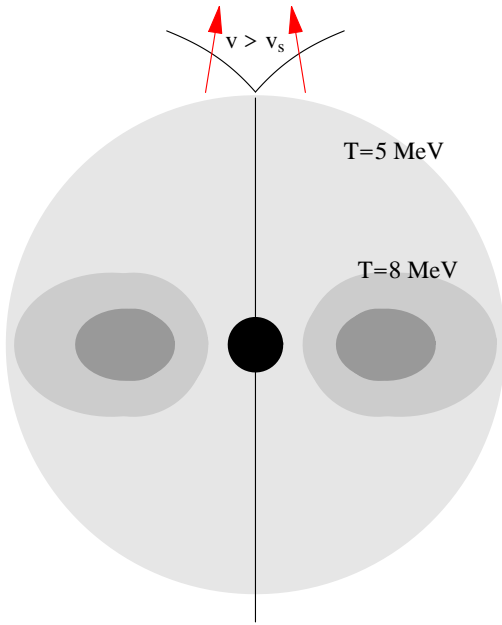} \\
            &
            \hbox{} \\
        \end{tabular}
    \end{center}
\label{fig. 1}
%\vskip 1in
\protect\caption
    {%
Cross-section of the accretion torus, showing the black hole
at the center. In the inner torus with darker shading, $\nu$-nucleon reactios dominate
the $\nu$ opacity. In the lighter gray area $e^+, e^-$ plasma dominates. The
edge of the lightly shaded area is our injection sphere. The curved lines at the top
form the boundary between subsonic and supersonic flow, and the arrows indicate flow
of $e^+, e^-$ plasma across this boundary. The temperatures are given in MeV.
}
\end {figure}

The figure shows a slice of the
system by a plane that is perpendicular to the plane of the accretion disk and
which contains the axis of the disk. The cross-sections of an inner torus
and that of an outer torus are shown with different shadings. In the inner one
we expect the neutrino opacity to be dominated by scattering and absorption
on nucleons, as it is in the simulations of refs \cite{nary} -\cite{janiuk}. Under the conditions of energy output that we consider, the innermost torus
may be $\nu$ opaque; it may be marginally transparent. The nucleon density
decreases steadily as one moves from the center to the edge of this torus\footnote{Most theories of the torus (or disk) proper, e.g. ref. \cite{nary}
assume relatively constant baryonic density as one moves up from the median plane
to the surface, whereas if one demands approximate hydrostatic equilibrium
in the presence of the (tidal) force needed to sustain rotation around a point
on the axis, but not at the center of the system, then the baryonic density must
decrease drastically as one approaches the boundary of the disk. This point was
made in the last section of ref. \cite{rfs}. It could be relevant to the model
being proposed here, since it provides a picture of a smooth transition between
the disk interior and the pair-plasma atmosphere being proposed here.}
In the surrounding toroidal
region, less shaded, as we move from the inside to the outside, the leptonic 
opacity becomes significant.
 In the more lightly shaded torus leptonic scattering takes over completely.
The very lightly shaded region is occupied by the plasma halo of our model. The $\nu$ opacities
are small enough so as not to change the $\nu$ fluxes and energy distributions
at the surface of the circumscribed sphere, our ``injection" sphere.

In estimates that we make in more detail in appendix A we find:

a) That for a neutrino-surface temperature of $5\,\,MeV$, located approximately
at the edge of the inner (dark) torus, the average temperature
of the overlying plasma is maintained at a temperature of about
3 MeV, with the heating provided by neutrino-electron and neutrino-positron
collisions, and the cooling being the usual $e^+ +e^- \rightarrow \nu_e+\bar \nu_e$
process. This heating mechanism was suggested by Woosley \cite{woosley} and also plays
a central role in the considerations of ref. \cite{asano}.

b) That the $\nu$ optical depth of our plasma sphere is approximately $.1$ under the
above conditions. 

In slightly more realistic estimates with the temperature varying with the
distance from the core we find that the region outside the large sphere contributes
only a small amount of additional optical depth; this was behind our choice of 50 km
as the radius. All of this begs the question of how the plasma got there in the first place.
We have made estimates based on seeding through the $\nu_e+\bar \nu_e\rightarrow e^+ +e^- $
process, followed by heating through the scattering processes, that indicate a
development time of the order of .01 sec.

We believe that there is an internal consistency in our assumptions,
but we cannot know that above picture will be sustained by future detailed
calculations of the complete accretion torus $\it cum$ transport calculation.
We do believe that the simulations to date lack an essential piece of physics
in the above speculations, namely the seeding of the far out region with
some pair plasma followed by plasma build-up through scattering processes.

Since our picture has at least an order of magnitude more energy deposited
by the neutrinos that leave the torus core than does the conventional model,
we have an incentive to look for acceleration mechanisms that can convert
some appreciable fraction of the energy to a relativistic jet. The purpose
of the present paper is to explore the hydrodynamics of the plasma configuration
described above. Our main result, which admittedly will feed in more assumptions,
will be that the temperature variation of the plasma as a function of $\theta$,
coming from the greater proximity of the torus for large $\theta$, at a given radius, can
at the same time enable the passage from subsonic to supersonic flow speeds
and concentrate the flow in a cone along the axis. 

The calculation assumes a steady flow of plasma across the surface of the large sphere, at
a subsonic speed, as the boundary condition for a steady flow solution of
the equations of fluid dynamics. However, for orientation, in the next two sections we 
consider examples of several kinds of flows, both time-independent 
and time dependent.
\section{Isotropic injection in the region close to the horizon}

As we shall consider only the temperature domain $T>2\,$MeV
in which the electrons are quite relativistic, we take the pressure to
be given by $P=b T^4/3$, where $b=(11 \pi^2/60)$ for an electron-positron-photon plasma,
with negligible contribution from nucleons. For the energy density, in some later examples, we add the contribution of nucleons of number density $n$, $\rho=b T^4+nM$, where $M$ is the
nucleon mass. We will remain in domains
in which the contribution of nucleons to the pressure is negligible.

The perfect fluid equations are \cite{weinberg},

 \begin{eqnarray}
{\partial P \over \partial x^\nu } g^{\mu \nu}+ {1 \over \sqrt g} {\partial \over
\partial x^\nu}[\sqrt g (P+\rho)U^\mu U^\nu]+
\nonumber\\
\Gamma_{\nu,\lambda}^\mu U^\nu U^\lambda 
(P+\rho)=h^\mu \, .
\label{eofm}
\end{eqnarray}

In the case in which $n\ne 0$ the equations (\ref{eofm}) are supplemented by the equation of baryon conservation. Here $h^\mu$ has non-vanishing components $h^0$, $h^r$ that are the respective
time rates of deposition of energy and momentum density in the plasma by neutrino 
annihilation, as seen by an observer at infinity. We comment more on these functions
below.
We define
\begin{eqnarray}
f(r)=-g_{00}(r)=(1-r_s/r)\,,
\end{eqnarray}
and introduce the variable $y$ of \cite{ft} and \cite{flam}
\begin {eqnarray}
y=[f(r)/(1-v^2)]^{1/2} \, ,
\end{eqnarray}
Henceforth we choose the Schwarzschild radius as the unit of distance, $r_s=1$.
Specializing to the
case of $n=0$, $P=\rho/3$, 
and substituting for the four-velocities in (\ref{eofm}), 
\begin{eqnarray}
U^0={f(r)^{-1/2} \over \sqrt{1-v^2}}~~,~~U^r={v f(r)^{1/2}\over \sqrt{1-v^2}}\,,
\label{4vs}
\end{eqnarray} 
we obtain, after much algebra,
\begin{eqnarray}
{\partial\over \partial r} ( v y^2 r^2 \rho) ={3 f\over 4}r^2 h^0+
{r^2\over 4}{\partial \over \partial t}\rho
-r^2{\partial \over \partial t}\Bigr ( {\rho \over 1-v^2} \Bigr )
\label{eofma}\, .
\end{eqnarray}
and
\begin{eqnarray}
{\partial \over \partial r}\log[\rho y^4]={1\over (1-v^2) \rho y^2}\Bigr [3(h^r-v f h^0) 
\nonumber\\
-v{\partial \rho\over \partial t}-4  \rho{\partial \over \partial t}
\Bigr ( {v \over 1-v^2} \Bigr )\Bigr ]\, .
\label{eofmb}
\end{eqnarray}

The neutrino annihilation sources that we simulate with $h^\mu$ in (\ref{eofma}) 
and (\ref{eofmb}) produce
pairs nearly at rest in the system of the observer at infinity, and at a rate that is 
essentially independent of the flow of the plasma into which they are injected. 
We implement this picture by adding an amount of energy density $\Delta \rho$,
measured in an inertial frame, at rest, at some point in the vicinity of the hole,
in time, $\Delta t$,  as measured by the observer at infinity. 
Then the right hand side of the eq (\ref{eofm}) for $\Delta(\partial_\mu T^{\mu 0})$ 
is,
\begin{eqnarray}
h^0=\Delta(\partial_\mu T^{\mu 0})=[U^{0}(v=0)]^2 \dot s =f^{-1} \dot s \, ,
\label{source}
\end{eqnarray}
where the four velocity $U^0$ is given by (\ref{4vs}), and $\dot s$ is the energy density deposition
rate, $\dot s \equiv (\Delta \rho /\Delta t)$.  We take $h^r=0$.

Since other authors use different source terms for the same equations, we elaborate
a little on our choice. If in the case $h^\mu=0$ we move everything in
(\ref{eofm}) except the time
derivative terms to the right hand side, then we describe the equations
as writing the change in time of $T^{00}$ and $T^{0r}$, respectively,
as a sum of changes due to compression and due to transport. When we introduce
the neutrinos, we add the additional changes 
to the energy-momentum tensor coming from their annihilation. This change,
over the infinitesimal $\Delta T$, is completely independent of the fluid
motion.

In contrast, in the case of the source terms used by Thompson, {\it et al.} \cite{bur},
as well as those used by ref. \cite{nobili}, and, we believe, by ref. \cite{aloy},
induce changes in $T^{00}$ and $T^{0r}$ that depend on the velocity of the
fluid. They are the correct equations when the source arises from radiative
transfer, but not appropriate for our case in which $\Delta T^{0,0}$
and $\Delta T^{0,r}=0$ are to be calculated by the observer at infinity.

It is nevertheless instructive to calculate the source function 
when we assume that energy density (and no momentum) is deposited at a given rate, $\dot q$, in the co-moving
system. The sources $h^0$ and $h^r$ are then,
\begin{eqnarray}
h^0
=\Bigr [{\Delta \rho \over \Delta t}\Bigr ][U^{0}(v)]^2={\dot q  y \over f } \, ,
\nonumber\\
\,
\nonumber\\
h^r=\Bigr [{\Delta \rho \over \Delta t }\Bigr ]U^{0}(v)U^r(v)={v \dot q   y}\, .
\label{altsources}
\end{eqnarray}
where we have now defined $\dot q$ to be the rate of energy density deposition as measured by
an observer in a co-moving inertial frame,

\begin{equation}
\dot q={\Delta \rho \over \Delta t \sqrt{1-v^2}\sqrt f}\, .
\end{equation}
Note that in this case the combination that provides the source term in
(\ref{eofmb}) vanishes, $h^r-v f h^0=0$. 

For the case of steady flow, the equations (\ref{altsources}) are equivalent to those of 
Thompson {\it et al} \cite{bur}, the latter specialized to the case of vanishing baryon density.
\footnote{The equations of ref. \cite {bur} derive in turn from those of ref. 
\cite{nobili}, further 
specialized to include only the first moment term in the energy absorption part of the
rediative transfer equation.} 

We begin discussion of the solutions with the case in which we set the 
right-hand sides of both (\ref{eofma})
and (\ref{eofmb}) equal to zero, looking for static solutions in the case of energy injection through a bounding surrface,
but with no volume injection of energy.
These equations now are exactly those of Flammang \cite{flam} (see also refs. \cite{ft} and 
ref. \cite{pac}) , specialized to the case
of the completely relativistic plasma with no nucleons present. Picking boundary conditions such that a constant temperature is maintained
at an injection radius $r_0>1$, and injecting at this point with velocity $v_0$ we find,
\begin{eqnarray}
v-v^3= (v_0-v_0^3) {r_0^2 f(r)\over r^2 f(r_0)}\, .
\label{vel}
\end{eqnarray}
In the
$P=\rho/3$ limit the magnitude of the energy density (i.e temperature) at the injection surface
scales out of the velocity determination.

Differentiating 
(\ref{vel}) we have, for all points above the injection radius,
\begin{eqnarray}
{dv \over dr} (1-3 v^2)={dv \over dr} (1-v^2/v_s^2)=
\nonumber\\
(v_0-v_0^3) {r_0^2 \over  f(r)}\Bigr({3 \over r^4}-{2 \over r^3}\Bigr)\, ,
\label{accel}
\end{eqnarray}
where $v_s$ is the speed of sound. Noting that the right hand side of
(\ref{accel}) vanishes only at the point $r=3/2$, we conclude that when 
the injection radius is outside this point the flow cannot make
a transition from subsonic to supersonic (or vice-versa); it is easy to
see that subsonic injection leads to a flow that decelerates, even in the absence
of gravity, as it 
moves upward, gradually filling all space with plasma and trading kinetic energy for 
internal energy. The velocity approaches zero at infinity. By contrast, supersonic injection  into a surface outside $r=3/2$ leads to an accelerating flow, with internal energy being traded for kinetic energy and an asymptotic velocity of c. This is the situation envisioned in models of
a gamma ray burst jet. (All of this, of course, is for the perfect $P=\rho/3$ plasma; in a real 
case, the plasma eventually cools to the point at which the $e^\pm$ are no longer relativistic.)

The options change when the plasma is launched from below the radius $r=3/2$. Now
there can a transition from subsonic (below) to supersonic (above) for upmoving
plasma if the injection velocity has exactly the right value (a function of $r_0$)
at the injection surface. 
If the injection velocity $v_0$ is greater than this critical value, there is no physical solution.
If the $v_0$ is less than the critical value, then the velocity will peak 
at $r=3/2$ and decrease thereafter. Fig. 2 shows
plots of the solutions to (\ref{vel}) both for a substantially subcritical injection 
speed and for a case that is subcritical by a tiny amount, the latter plot showing the
non-avoided crossing of two roots that is the key to the transition.
\begin {figure}[ht]
  \begin{center}
     \epsfxsize 2.75in
        \begin{tabular}{rc}
           \vbox{\hbox{
$\displaystyle{ \, { } }$
               \hskip -0.1in \null} %\vskip 0.2in
} &
            \epsfbox{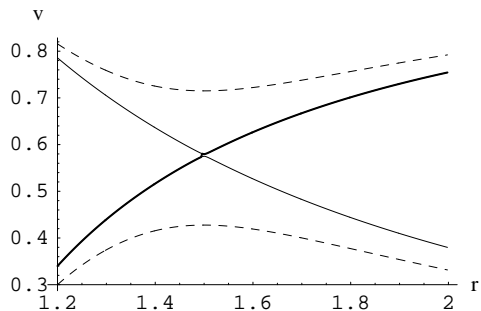} \\
            &
            \hbox{} \\
        \end{tabular}
   \end{center}
\label{fig. 2}
\vskip 1in
\protect\caption
   {%
Velocity as a function of radius for two different values of injection velocity,
where the injection surface is at $r=1.2$. The solid curves show the two relevant
solutions to the cubic equation (\ref{vel}) for the case of injection velocity
$v(1.2)=.34$, a value very slightly less than the critical value. The heavy
overlay shows the form of the solution that (when the injection is exactly at the
critical velocity) connects through the sonic point. The dashed curves are for
injection at a slightly lower velocity, $v(1.2)=.3$; the lower dashed curve is
a physical possibility in which the flow remains subsonic in the entire space.
}
\end{figure}

All of the above analysis, and in particular the plots of fig. 2, are essentially
the same as presented by Flammang \cite{flam}. In the application
to accretion flows in this reference and others, where $v<0$, the transition is from
subsonic (above) to supersonic (below), 
and the subcritical solutions represented by the (dashed) curves in fig. 2
can be dismissed as unphysical \footnote{In the case of accretion flow, 
where the velocity in fig. 2 is now a downward velocity, the physical curve
is the one with negative slope. }. 
In our problem, though, we shall argue that the
solution analogous to the lower dashed curve in fig. 1 could be the one chosen,
once we put in the source terms, $h^0, h^r$.

For the moment, we keep the source turned off, again taking boundary conditions
for injection at a surface, but now turning on the injection at $t=0$ and solving
the time dependent equations, to see if there is a stable development of 
steady flow at sufficiently long times. Our computational abilities
are not sufficient to deal with the sonic transition for the time
dependent case, but it is illuminating to look at the time behavior both in a 
completely subsonic case and in a completely supersonic case.

We begin with the case of equations (\ref{eofma}), (\ref{eofmb})
with the source terms set equal to zero, and consider the two generic cases of injection
through a surface located at some $r_0> 3/2$. 
Starting at time $t=0$ with nearly empty space above the injection surface, we maintain a constant
energy density and inflow velocity on the surface. To avoid a singular 
space derivative at the boundary we took an initial energy density profile $\exp -a(r-1)$ instead of absolutely empty 
space for $r>r_0$ at $t=0$ as the initial condition. In figs 2, 3 we show time behavior for the supersonic case 
 $r_0=7 $, $a=1.5$, and $v_0=.7$.

We see that a steady flow solution does indeed establish itself in short order
, in a continually expanding region, a flow in which the initial kinetic
energy, determined by the injection velocity $v_0$, is being converted to internal energy.
The profiles one sees at the rear of the solid figures show steady flow conditions
with the velocity plot conforming to the solution of (\ref{vel}).

\begin {figure}[ht]
  \begin{center}
     \epsfxsize 2.75in
        \begin{tabular}{rc}
           \vbox{\hbox{
$\displaystyle{ \, { } }$
               \hskip -0.1in \null} %\vskip 0.2in
} &
            \epsfbox{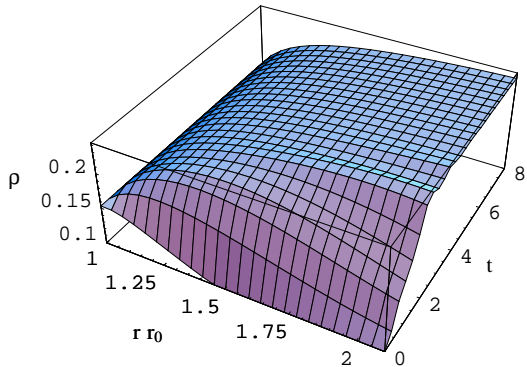} \\
            &
            \hbox{} \\
        \end{tabular}
   \end{center}
\label{fig.3}
\vskip 1in
\protect\caption
   {%
Energy density, $\rho$ (or $T^4$), in arbitrary units, for the case of 
steady subsonic injection at
a distance $r_0=7r_s$, with injection velocity 
$v(1)=.15$. The injection is turned on at
$t=0$. The plot shows the rapid evolution into steady flow in the region $1<r/r_0<2$,
and the conversion of kinetic energy to internal energy
in the decelerating flow. The flow moves from left to right in the figure.
}
\end{figure}

\begin {figure}[ht]
    \begin{center}
       \epsfxsize 2.75in
        \begin{tabular}{rc}
           \vbox{\hbox{
$\displaystyle{ \, { } }$
               \hskip -0.1in \null} %\vskip 0.2in
} &
            \epsfbox{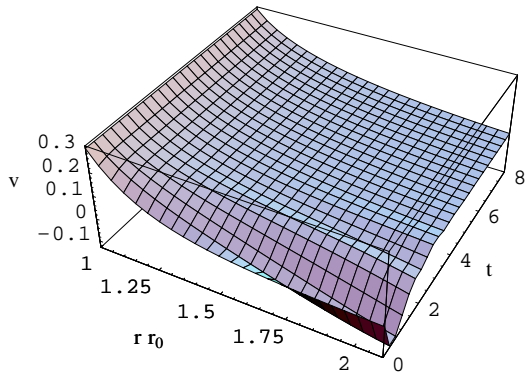} \\
            &
            \hbox{} \\
        \end{tabular}
    \end{center}
\label{fig.4}
%\vskip 1in
\protect\caption
    {%
Outward velocity for the case of steady subsonic injection, under the same conditions
as used in fig.3, showing the slowing of the flow as it moves upwards. 
}
\end {figure}

For comparison, in figs. 5,6 we show the case of supersonic injection, with initial velocity
$v=.7$. Again, the steady flow establishes itself quickly. We see the matter speeding
up rather than slowing down, with internal energy being traded for kinetic energy, as in the standard
scenario in the literature.

\begin {figure}[ht]
    \begin{center}
       \epsfxsize 2.75in
        \begin{tabular}{rc}
           \vbox{\hbox{
$\displaystyle{ \, { } }$
               \hskip -0.1in \null} %\vskip 0.2in
} &
            \epsfbox{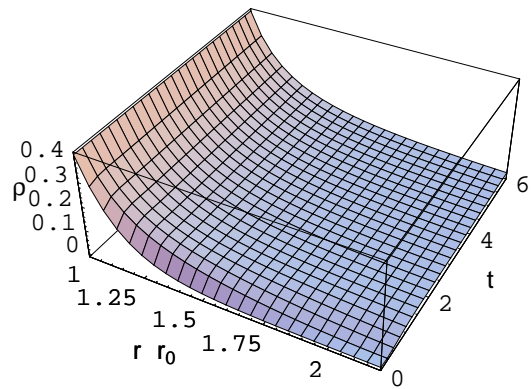} \\
            &
            \hbox{} \\
        \end{tabular}
    \end{center}
\label{fig. 5}
%\vskip 1in
\protect\caption
    {%
Energy density for the case of supersonic injection with injection velocity of .7c .
}
\end {figure}

\begin {figure}[ht]
    \begin{center}
       \epsfxsize 2.75in
        \begin{tabular}{rc}
           \vbox{\hbox{
$\displaystyle{ \, { } }$
               \hskip -0.1in \null} %\vskip 0.2in
} &
            \epsfbox{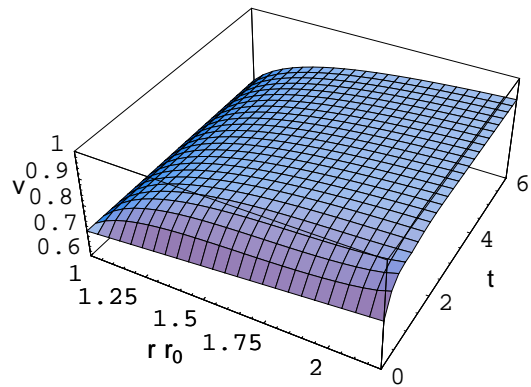} \\
            &
            \hbox{} \\
        \end{tabular}
    \end{center}
\label{fig. 6}
%\vskip 1in
\protect\caption
    {%
Velocity evolution for the case of supersonic injection. Taken together with fig. 5,
it shows the upward acceleration through conversion of internal energy to kinetic 
energy.
}
\end {figure}

We have used the same codes to check on the stability of solutions against radial perturbations,
by taking an initial distribution that is a fit to the steady flow limit, adding localized perturbations
at different distances, and confirming rapid relaxation back to the steady flow solution, both for the
cases of subsonic injection and for supersonic injection. 

The above discussion was given in order to illuminate the problem that we set out
to solve, but that problem was defined by a given rate of energy-momentum deposition
in the entire space rather than by injection rates through a boundary surface.
Turning to the case $h^0,h^r\ne 0$ in (\ref{eofma}), (\ref{eofmb}), still for the
case of steady flow, we now ask for solutions for the whole space, from the horizon
at $r=1$ to $r=\infty$. 

We begin with an example that can be reduced to a cubic
equation by quadrature (although it is both artificial and doesn't
address the case $h^r=0$, $h^0\ne 0$ in which we are the most interested). This is the case
in which
 $h^r-v f^{-1}h^0=0$, so that the right-hand side of (\ref{eofmb}) is zero, and we can use
 $\rho=C_1^{-1} y^{-4}$ to eliminate $\rho$ in (\ref{eofma}). This choice corresponds to injecting pure energy density
and no momentum density as seen from the frame moving with the plasma. Thus it doesn't
satisfy the ground rules of being externally provided in the observer-at-infinity
frame, and not dependent on the details of the flow itself.
But the example gives a neat illustration of the behavior of solutions when we have volume
injection rather than injection through a lower boundary surface.

Following ref. \cite{aloy} we chose a power-law radial dependence, 
\begin{eqnarray}
h^0(r)=r^{-5}\,,
\nonumber\\
h^r-v f^{-1}h^0=0 \, ,
\label{sources2}
\end{eqnarray}
where in the first of these equations, for the rate of energy deposition $h_0$, we use
used the exponent suggested by  ref.\cite{aloy}. In this reference, the dependence of the
source strength on position in the region near the black hole is taken as $(r {\cos(\theta}))^{-5}$
(in some limited angular region) where $\theta$ is the polar angle. The polar angle dependence of the source is important,
and we treat some aspects of it in the next section, after exploring examples of purely
radial flow.

Integrating (\ref{eofma}) for the source choice (\ref{sources2}) we obtain,
\begin{eqnarray}
v-v^3=  C_1 (1-1/r){(r_1^{-2}-r^{-2})\over r^2 }\,,
\label{vel2}
\end{eqnarray}
where the integration constant $r_1$ is the stagnation radius.

For the plasma to escape at all there must be a sonic point $r_0<r_1$ where, as we move up, we
go through the value $v(r_0)=-1/\sqrt 3$. All energy injected below the stagnation
radius $r_0$ goes down the hole.
By trial and error, we can adjust the two integration constants $C_1$ and $r_1$ in order to
find the single solution that passes through the two sonic surfaces that now present themselves,
that is, the transitions at $v=\pm 3^{-1/2}$. In fig. 7 we show basically this solution, but with
the integration constants detuned very slightly so that one sees the normal avoidance of crossing of
the different solutions of the cubic.  

\begin {figure}[ht]
  \begin{center}
     \epsfxsize 2.75in
        \begin{tabular}{rc}
           \vbox{\hbox{
$\displaystyle{ \, { } }$
               \hskip -0.1in \null} %\vskip 0.2in
} &
            \epsfbox{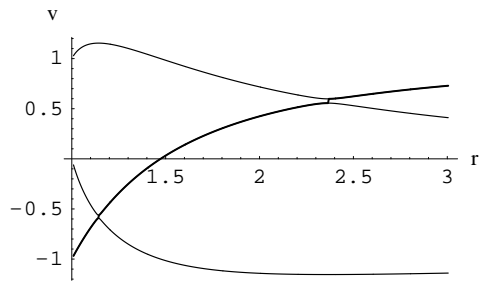} \\
            &
            \hbox{} \\
        \end{tabular}
   \end{center}
\label{fig. 7}
\vskip 1in
\protect\caption
   {%
The solution to (\ref{vel2}) for the unique parameters that connect the supersonic
ingoing flow at the horizon to supersonic outgoing flow at large distances. The
red curve gives the critical path that can be achieved when the parameters are exactly tuned.
}
\end{figure} 
In this solution approximately one half of the energy deposited
goes down the hole. There is also a one parameter group of solutions that go through the first sonic
surface, and pass through a $v=0$ surface, but which never go supersonic, with the velocities
decreasing at large heights as described previously. 

In fig.8 we show a solution for the same equations, but with slightly different input
parameters $C_1$ and $r_1$ which lead to a solution where some of the flow does escape,
but in which the flow remains subsonic and ultimately slows.

\begin {figure}[ht]
  \begin{center}
     \epsfxsize 2.75in
        \begin{tabular}{rc}
           \vbox{\hbox{
$\displaystyle{ \, { } }$
               \hskip -0.1in \null} %\vskip 0.2in
} &
            \epsfbox{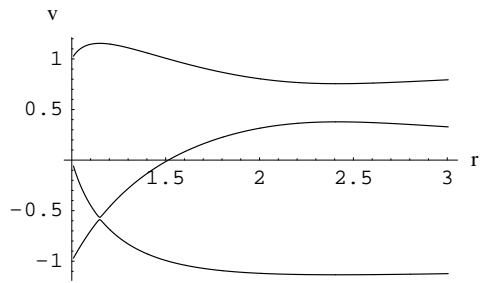} \\
            &
            \hbox{} \\
        \end{tabular}
   \end{center}
\label{fig. 8}
\vskip 1in
\protect\caption
   {%
The red curve shows the one parameter set of solutions to (\ref{vel2}) in which flow does escape to infinity
but remains subsonic. The input parameters $C_1$ and $r_1$ are only slightly different
from those used in fig.7, with the stagnation point $r_1$ being slightly farther out.}
\end{figure} 

As we noted above, to get the second of equations (\ref{sources2})we assumed that 
energy density, and no momentum, is deposited  at given rate in the co-moving system, rather than
in the observer-at-infinity frame, leading to (\ref{altsources}). This is consistent
with the assumptions of ref.\cite{aloy}, as we can see from reverse engineering
eq. 2 of this reference, which discusses the relationship of the source term
in the two frames.\footnote{We note also that $h^r-v f^{-1}h^0=0$ is required to obtain the equations
used in ref. \cite{bur} in a rather similar application} 
 In the present case, the assumption allows the differential
equation for the velocity field to be reduced to a simple cubic algebraic equation,
shown below. However with this choice the rate of
deposition of energy-momentum in the plasma depends through $v$ on the motion of the plasma,
which it should not, in the physical case. 

Next we do the parallel analysis for the case of $h^0(r)=r^{-5}f(r)^{-1}$, $h^r(r)=0$, that is, pure energy deposition
as viewed by the observer at infinity. We numerically solve the coupled differential equations, in the three 
regions: horizon to first sonic point; first sonic point to second sonic point; second sonic point
to infinity. Adjusting the initial conditions in each segment, by trial and error, leads to a plot that is so similar to the heavy curve in fig. 7 that we do not need to plot it. The $v=0$ point is at  $r_1=1.48$,
and we find that 44\% of the energy deposited escapes; the remainder goes
down the hole. This solution that becomes supersonic
is unique, but again there is a one-parameter family of perfectly acceptable solutions that 
remain subsonic. This leads us to our first important conclusion: {\it Specification of steady 
energy and momentum deposition rates throughout the whole space does not determine
a unique steady flow solution}.

Presumably, which, if any, of the steady flow solutions is obtained a long time
after some turn-on of the sources will be determined by the solution of a time
dependent problem, and the solution could depend delicately 
on pre-existing conditions such as the configuration of plasma present prior
to the turn-on or it could depend on the profile of the turning on function.
We do not have the resources to explore the establishment of the sonic points
through solution of the
time dependent equations, except in a fragmentary way. However, in fig. 9 we show the results of a time dependent 
solution where we have
both a lower bounding surface with fixed subsonic outward velocity on the surface, and 
volume injection of energy defined by $h^0=r^{-5}, h^r=0$, where
the injection is turned on at $t=0$. In the profile at the rear
of the 3D plot we see the steady flow solution that develops, which
remains subsonic. This calculation is supportive of the idea
that stable, totally subsonic flows are a possible outcome.

\begin {figure}[ht]
  \begin{center}
     \epsfxsize 2.75in
        \begin{tabular}{rc}
           \vbox{\hbox{
$\displaystyle{ \, { } }$
               \hskip -0.1in \null} %\vskip 0.2in
} &
            \epsfbox{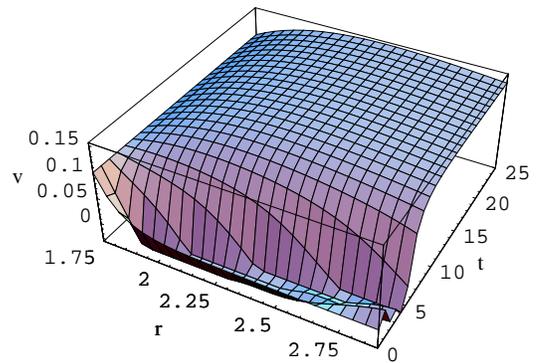} \\
            &
            \hbox{} \\
        \end{tabular}
   \end{center}
\label{fig. 9}
\vskip 1in
\protect\caption
   {%
An example of a time dependent solution of (\ref{eofma}),(\ref{eofmb}), with
source term as described in text, above a bounding surface at $r=1.75$,
where the injection velocity is $.1$, showing, 
in the rear profile of the figure, a steady 
flow that remains subsonic over all space above the boundary surface, with the velocity
reaching a maximum at about $r=2.3$ }
\end{figure} 

Finally, going back to the steady flow equations, we have added  massive nucleons, in small enough percentages
 so that they contribute significantly to the energy density 
but negligibly to the pressure. Here, in the supersonic case, there is an interplay between the 
injection radius and initial nucleon contamination that can limit the asymptotic value of
$\Gamma$ to values much lower than the range $\sim 100$ that is desired. A qualitative
statement is that when the energy density in the $e^\pm$ gas is fifty times 
the rest energy density in the nucleons, the Lorentz factor
will approach $70$ asymptotically, for the case of injection at 7 times the Schwarzschild
radius and at an initial speed 5\% greater than the sound speed. If we take
the $e^\pm$ plasma to have a temperature of $4.5 MeV$, at injection, this corresponds
to a nucleon mass density $\rho_N\approx  10^6 g c^{-3}$. If we operate with significantly less than this amount of contamination, the influence of gravity on the flow is negligible.

The results of this section seem to open the door to a modification
of the standard picture of the very inner core of a $\nu$ driven jet, in which
in the innermost region, where energy deposition into the plasma is from the
annihilation source, the plasma may be decelarating.
It is true that the authors of ref \cite{aloy}, in a time dependent simulation,
find a rapid evolution to a steady flow solution that makes both sonic transitions 
without incident. This may be completely due to the angular dependence
built into the authors' $z^-5$ dependence of the source term,
or it could be because the time-dependent simulation chose a steady
flow limit embodying the sonic transitions. The plot of $\Gamma$
shown in fig. 4  of ref. \cite{aloy} is on such a large scale of $r$ as to make the
region of neither transition visible in detail.

Moreover, even though we use a function similar to that of these authors
for the rate of energy deposition
(as seen by the observer at $\infty$),
it appears that this function enters into our equations in a different way. We note
equation (2) in ref \cite{aloy}, relating the rate of energy deposit $\dot q$ as 
seen by the observer at $\infty$ to the rate $\dot q'$ as seen in the comoving system,
\begin{eqnarray}
\dot q'={\dot q \over f^{1/2} [ \gamma (\Gamma^2-1)+1]}\, ,
\end{eqnarray}
where $\gamma=4/3$ for our case. This equation correctly describes the relation
for the case in which the
external source is taken to have $T'^{r0}=0$ in the comoving system,
that is, to be at rest in this system. As mentioned above, this makes the rate of momentum
deposit by the annihilation process dependent on the flow velocity of the plasma. 
Our simulations confirm that the assumption gives a bias toward
escape in the following sense: even though we find that specifying the source
{\it does not} uniquely determine the steady flow, if we compare 
the solutions which do pass through both sonic points, and accelerate
all the way to $r=\infty$, a significantly smaller fraction of the deposited energy is captured
in the case in which the pure energy is added in the comoving system, as opposed to the
observer-at-infinity system.

In any case, if we need to go to a time dependent solution to pick which
of the different steady state flows eventuates (for a fixed energy density injection profile), then
we would like to know what external factors affect the answer. It might be helpful in
understanding the close-in physics to know what the large relativistic hydrodynamics codes
give for the detailed flow near the hole, the location and shape (for non
spherical solutions) of both sonic surfaces, as a function, say of a turn-on profile for the source
or the distribution of initial contamination by nucleons.

In any case, in view of our results on the physical plausibility
of the flows that remain subsonic, we assume for the remainder of this paper
that the (conventional) annihilation engine indeed produces a plasma that slows
(but that does not cool much) as it moves outwards, filling the close-in
regions. This plasma can be helpful
in providing a seed for the plasma halo that we discussed in the introduction,
but the energetics of this halo will be sustained largely by $\nu$ scattering off of the
plasma particles rather
than by $\nu,\bar \nu$ annihilation processes.

\section{Anisotropic injection}
The considerations of the previous section were confined to a region extending to
a distance of two or three times the horizon radius. Now we go to a distance of roughly
ten times the horizon radius, as sketched in fig. 1, and discuss the dynamics
of a nonisotropic plasma, with higher densities at higher angles since
the higher angle region at a given distance has greater proximity to the 
neutrino-emitting torus. We shall demonstrate a path through the sonic transition
that depends exactly on this angular dependence. Gravity is nearly irrelevent
at the distances in question. Nor will we include a source term, $h^\mu$
as in (\ref{eofm}), although there will be heating from $\nu$ scattering
and cooling, the latter from $e^+, e^-$ annihilation, roughly in balance.

Here we consider only the steady flow case. But the time dependent solutions
presented for the isotropic case give us confidence that steady flow 
solutions will establish themselves in a growing volume above the injection
point as time goes on. We begin again with (\ref{eofm}), but now take
$\rho (r_0,\theta)={\rm const. }(1+.7 \sin \theta)$ on the inner boundary.
This corresponds to a temperature that is roughly 10\% greater at the equator than
at the pole. 
In figs. 8,9 we show an example of the resulting flow as determined from a solution in a
region $\theta_0< \theta< \pi/2,~ r_0<r<r_1$. The solution in the region 
$\pi/2<\theta <\pi$, would be obtained by up-down symmetry. The derivatives $\partial /
\partial \theta$ of all quantities vanish for $\theta=\pi/2$.
In the example shown we took,
$\theta_0=.12$ 

\begin {figure}[ht]
    \begin{center}
       \epsfxsize 2.75in
        \begin{tabular}{rc}
           \vbox{\hbox{
$\displaystyle{ \, { } }$
               \hskip -0.1in \null} %\vskip 0.2in
} &
            \epsfbox{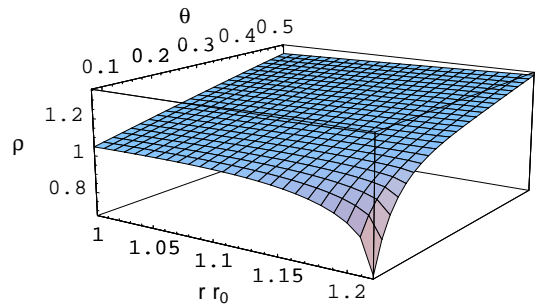} \\
            &
            \hbox{} \\
        \end{tabular}
    \end{center}
\label{fig. 10}
%\vskip 1in
\protect\caption
    {%
Density contours in the steady flow solution for the non-isotropic case, showing the
increases of magnitude in both components of the velocity as we approach the sonic
singularity of the near corner of the plot. This point is just one point on the sonic
surface that we explore by repeatedly using different values of ($r_1, \theta_1$) (the corner
coordinate), each of the pairs being chosen to put the maximum velocity point on the corner.
}
\end {figure}

\begin {figure}[ht]
    \begin{center}
       \epsfxsize 2.75in
        \begin{tabular}{rc}
           \vbox{\hbox{
$\displaystyle{ \, { } }$
               \hskip -0.1in \null} %\vskip 0.2in
} &
            \epsfbox{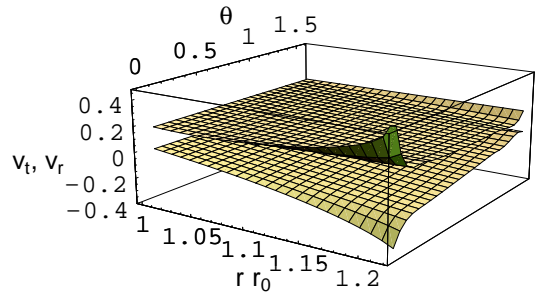} \\
            &
            \hbox{} \\
        \end{tabular}
    \end{center}
\label{fig. 11}
%\vskip 1in
\protect\caption
    {%
Radial and tangential velocity contours in the steady flow solution for the non-isotropic case.
The upper sheet shows the radial velocity, and the lower the tangential velocity, with
negative values since it is inwards. At the near corner of the plot, $r_1,\theta_1$}
\end {figure}

We designate the nearest corner on the lower plane of the bounding cubes in figs. 9,10 as the
the point $r_1,\theta_1$.
Approaching this corner,
we see a rapid rise in $v_r$ as well as the increase in the magnitude of $v_t=r {d \theta\over dt}$ (which is negative). In the
example shown, the resultant speed at this corner is very close to the
speed of sound ($1/\sqrt 3$). Repeating this calculation
with different input values of $\theta_0$ allows us to plot out the envelope 
of the supersonic region.

Going back to fig.1 we show the part of the sonic boundary 
that is nearest to the injection surface as a pair of curves sprouting out of the top of
the injection surface.
As we noted, the velocity field has a gradient approaching infinity
as we approach the surface from the outside, with the derivative of the energy
density approaching (negative) infinity as well, as it must to conserve energy.
On the inside of the surface, as we we move upwards we then will have
the continuation of upward (but now decreasing) acceleration to velocities
approaching the speed of light.

In the previous section we investigated the sonic transition in isotropic
flows. These transitions were driven by a combination of gravitational
effects and the radial dependence of the source terms; in the absence
of both there would have been no sonic transition.

In the present section we found that a non-isotropic flow can make the transition
on its own, that is to say, in the absence of gravitational effects or infusion
of energy from the outside. This encourages us to take the model described
in the introduction seriously. In this model the plasma halo moves outward
with small velocities in a region in which $\nu, \bar \nu$ annihilation
no longer provides a big energy source, the heating being instead through $\nu, e^\pm$
collisions. The anisotropic temperature created by this heating mechanism
gives the needed precondition for a more or less conical sonic surface to develop,
and for accelerating supersonic flow to prevail in the interior of the cone,
was shown schematically in fig. 1. \footnote{The authors of ref.\cite{mcl},
after analyzing energy deposition by neutrinos, found
higher temperatures off-axis than along the axis and suggested that this might 
strengthen the jet}

\section{Discussion}

We search for a mechanism that can accelerate plasma to supersonic speeds,
and thence to
a Lorentz factor of up to $\Gamma=100$. The second stage appears to be easier
than the first; our solutions for a spherically symmetrical case with
isotropic supersonic injection are similar in radial structure
to the models
of refs  \cite{le} \cite{aloy}, where the injection is confined to a cone (and lateral spreading is
restricted by pressure from the medium.) Even in these models with a confined jet
it should be the case that the plasma configuration is more spherical than jet-like
in the innermost region, $r<({\rm a~few}) r_s$. Furthermore, once we enter the domain $v\approx c$ 
we recover the scaling relation $\Gamma (r) \sim r$ found in the analytic results of ref \cite{le}, and 
confirmed approximately in the numerical results of ref. \cite{aloy}. However,
in the spherically symmetrical case with subsonic injection, the
flow slows as it moves upwards, stagnating at $r=\infty$. 

The situation changes when we introduce a dependence of the
injection energy density on the polar angle $\theta$. 
Distinguishing two cases, roughly, we can consider either an injection
temperature that is greater at the equator, the case that corresponds to
our qualitative picture of how the halo forms, or an injection temperature that is greater at the 
poles. In the first case we indeed find a flow that accelerates in the region
near the axis, with the resultant of radial and tangential velocities reaching the 
sound speed as we approach, from the outside, a hornlike surface (that is, cone-like with expanding angle), centered on the axis. 

The hydrodynamical equations are singular as we approach this surface.
For the boundary condition for the region on the inner side of this surface we take the 
values of energy density and directions of velocity from the previous
calculation, boosting the magnitude of the velocity to a value barely above the
speed of sound, providing a new boundary condition for the acceleration up
to large $\Gamma$ in the inside of the sonic cone.

In the case in which the subsonic injection density is greater toward the equator than toward the
poles, we find that the velocity decreases, going upward, for all values of $\theta$. 
We conclude that having the injection temperature higher toward the equator
is essential in forming the jet in the vicinity of the polar axis.

All of our simulations in the non-isotropic case were for a completely 
relativistic plasma, $P=\rho/3$. But we believe that the numerical
experiments that we mentioned in sec. 2 for a plasma with nucleon loading
support the conclusion that in the domains of density 
of $\rho< 10^6 g c^{-3}$ the qualitative behavior is not changed by the loading.

It would be interesting to see the detailed results of 
large time dependent codes that have been applied to
the study of the large scale structure of the jets, for example, GENESIS, 
applied locally in the region near the black hole under a number of assumptions
for the initial conditions that prevail,
in order to better understand the universe of possibilities in this region

Comparing to the model of ref. \cite{nary}, which assumed a disk with a sharp boundary in which the density is
independent of height above the median plane, 
we find that even if we begin with this system, and a (near) vacuum on top, a 
pair-plasma will be created in the region above the $\nu$-surface, at a
temperature somewhat less than that of the $\nu$-surface. The seed would be from
$\bar \nu +\nu\rightarrow e^++e^-$ interactions, either produced locally or
drifted in from below, but the heating to final
temperatures is dominantly from $\nu-e^\pm$ collisions. Estimates leading to these conclusions are given in the appendix.

This hypothetical time-dependent scenario deserves to be addressed with a real transport calculation,
which we have not done. However, we had previously looked \cite{rfs} at a steady flow solution for the
the vertical structure of an atmosphere with a sharp surface, with accretion rates such
that the optical thickness (to $\nu$'s) is large, where nuclear opacity, relativistic
opacity, and vertical hydrostatic equilibrium were taken into account. We found that
the nuclear densities must decrease steadily as we go up, and end at a very small value at the surface, a finding that coexists well with the assumptions of the present paper.

That said, we return to reasons why the present work is not
in detail a realistic model of the jet formation:

1) We may have been over-optimistic in taking small nucleon densities ($\rho<10^8$)
in the region of larger angles $\theta$. But this is the problem that we could
formulate and solve. Of course, as we approach $\theta=\pi/2$ the densities
get really large. However, had we constrained to a cone bounded by $\theta_{max}$
we would have had to put an arbitrary boundary condition on the surface of the cone,
whereas in the solution that has the full $4\pi$ solid angle, there is no such
boundary, only the up-down symmetry that guarantees that the derivatives $\partial/\partial \theta$
vanish at $\theta =\pi/2$.

2) Our injection sphere had an arbitrary radius, so the curve which in fig. 1
shows the sonic surface is thus arbitrarily situated as well. (Of course, our assumption
that the injection velocity is radial is arbitary as well.) Suppose that we
had taken the radius to be a little bit smaller, and then calculated the temperatures
and velocities on the first surface, as a new boundary condition. Then transverse
velocities would no longer vanish on the injection surface. We would then have 
found the horn-shaped sonic surface to start lower, in fact lower than the 
injection radius for a $\theta=0$. Our actual choice of $7 r_s (\approx 50$km, 
for a three solar mass BH) was motivated by a compromise between these considerations and 
the need both to get outside of regions of high nucleon density and regions
in which the back reaction from neutrino transport would be a large perturbation
on the perfect fluid dynamics.

For the moment ignoring these drawbacks, we ask how the energy injected in the outward beam by the mechanism of the present
paper could compare with that in, e.g., the models of ref.\cite{aloy}. In a representative
example of the latter models, as the jet passes our surface at $50$ km., it carries
something like $10^{50}-10^{51}$ ergs/sec. across this boundary. The temperature 
of the plasma in the jet is already well less than $1$ MeV at this point. By contrast,
in our guesswork example, the temperature just above the injection radius, but at somewhat
higher angle $\theta$ is of the order $3$ MeV. As is clear from fig. 6, as we move to
smaller values of theta and approach the sonic surface, the energy density (or $T^4$)
drops rapidly, and it will continue to drop as we move away from the sonic surface
in the inside supersonic region. However the plasma energy density inside the
lower part of the sonic ``horn" still should be greater in this model than in the
model of ref. \cite{aloy}

In this
more muscular jet, the tolerance for baryonic contamination would likewise be much greater.
From the standpoint of total energetics, the new scenario uses the ability
of the plasma halo, which has greater volume than the farther-in annihilation
region of ref \cite{aloy}, to capture energy from a larger fraction of the
total $\nu$ output of the torus and deliver some fraction to the jet.
Some more detailed considerations on the energetics of heating mechanisms are
given in the appendix.

The core results of this paper remain the solutions of the hydrodynamical equations
in the perfect fluid domain. 
Some of the qualitative properties of these solutions have not surfaced in other results
of numerical computations known to us. At the least, it would be interesting 
to test the more comprehensive codes against our results, or against similar
computational fragments based on simplified systems. As we learned during the
generations of supernova calculations, once ``all the important physics" is put into
calculations of two different groups, it is certain that the calculations are not
exactly the same as each other, and comparison of the results is problematical.
\section{Acknowledgements}
During the course of this work I have profited greatly from communications with
Thomas Janka. The work was supported in part by NSF grant PHY-0455918. 
 
\section{Appendix}

Here we discuss the energetics of the configuration sketched in fig. 1, which provided
some of the motivation for the considerations of sec. 3. 
The figure shows a slice of the
system by a plane that is perpendicular to the plane of the accretion disk and
which contains the axis of the disk. The cross-sections of an inner torus
and that of an outer torus are shown with different shadings. In the inner one
we expect the neutrino opacity to be dominated by scattering and absorption
on nucleons, as it is in the simulations of refs \cite{nary} -\cite{janiuk}. The nucleon density
decreases steadily as one moves from the center to the edge of this torus; at the edge
completely leptonic opacity is becoming important. In the surrounding toroidal
region, less shaded, as we move from the inside to the outside the leptonic 
opacity becomes dominant. For our choice of parameters the $\nu$ free path
in this region is about $30\,$km., a little too 
long to signal $\nu$ trapping, but short enough to be relevant. For a case with
slightly higher temperature we would have actual trapping. Thus we picture the
$\nu$-surface as lying somewhere between the inside and outside of the region
with medium shading.
Outside of this region, in the very lightly shaded region, the $\nu$ opacities
are small enough so as not to change the $\nu$ fluxes and energy distributions
at the surface of the circumscribed sphere, our ``injection" sphere.

A shell of plasma at the edge of the lightly shaded
region, for example at the point B labeled ``$T=2.9$ MeV" is cooled by thermal $\nu$ emission
and heated by scattering of neutrinos from below. The latter have an energy distribution
corresponding a $\nu$-surface at $T=5$ MeV, but their number density is reduced at the point B
by a factor $\lambda$=({\it solid angle subtended at B by $\nu$-surface})$/4 \pi\approx .1$. 
The temperature at point B will be set by balancing the heating and cooling rates.

To estimate the rate of cooling we begin with the total cross-section
for the process $ e^- +e^+\rightarrow \nu_e+\bar \nu_e$, in the completely
relativistic limit, with the four-momenta
of the electrons denoted by $p,q$, and taking $\sin^2(\theta _W)=.25$,

\begin{equation}
\sigma_{e^+, e^-}= {5 G_F^2 \over 12 \pi} p_\mu q^\mu\, .
\end{equation}
Defining, $T_P$ as the temperature of the plasma at point B, and $T_S$ as the
temperature of the neutrino-surface, we write the cooling rate, per unit volume, at B
as,

\begin{eqnarray}
&\dot q_{\rm cool}= 4 (2 \pi)^{-6} \int d^3p\,d^3 q f(E_p,T_B) f(E_q,T_B) (E_p+E_q)
\nonumber\\
&\times \sigma_{e^+, e_-}=.35 G_F^2 T_B^9
\nonumber\\
&\approx 3.8 \times10^{33} T_{11}^9 {\rm ergs\,(cm)^{-3}\, sec^{-1}},
\label{cooling}
\end{eqnarray}
where f(E,T) is the Fermi distribution with zero chemical potential.

Next we estimate the net heating rate at B beginning from the sum of elastic cross-sections,
\begin{eqnarray}
\sigma_T=\sum_{i=e,\nu,\tau} ( \sigma_{\nu_i, e^-}+\sigma_{\bar \nu_i, e^-}+
\sigma_{\nu_i, e^+}+\sigma_{\bar \nu_i, e^+})
\nonumber\\
= {14 G_F^2 \over 3 \pi} p_\mu q^\mu \,,~~~~~~~~~~~~~~~~~~~~~~~
\nonumber\\
\end{eqnarray}
where the sum is over the three $\nu$ flavors, and
where $p$ and ${ q}$ are now the respective momenta of the incident neutrino and electron.
We estimate the rate of energy delivery as,
\begin{eqnarray}
\dot q_{\rm heat}=  \lambda(2 \pi)^{-6} \int d^3p\,d^3 q (E_p-E_q) f(E_p,T_B)
\nonumber\\
\times  f(E_q,T_S)\sigma_T
=3.4 \,\lambda \, G_F^2 T_B^9\, ,~~~~~~~~~~~
\label{heat}
\end{eqnarray}
where the initial factor of $2$ is from the sum over $e^\pm$ spins (the cross-sections
quoted are averaged over initial spins. The expression (\ref{heat}) vanishes 
when $T_S =T_B$, as it should. To obtain the last equality in (\ref{heat})
we took $T_S={5\over 2.9}T_B$. For this choice, $T_S=5 MeV\,\,T_B=2.9$, with
$\lambda\approx .1$, comparing with (\ref{cooling}), we see that the heating from
scattering is roughly in balance with cooling from $e^+, e^- $ annihilation.

We check our assertion that the heating through the scattering process is
greater than that from the conventional neutrino annihilation process. The
rate for the latter is, from(\ref{heat}),
\begin{eqnarray}
\dot q_{\rm ann}=\lambda^2  4\eta 2\sigma_{e^+, e^-}(2 \pi)^{-6} \int d^3p\,d^3 q f(E_p,T_S) 
f(E_q,T_S)
\nonumber\\
\times(E_p+E_q)\sigma_{e^+,e^-}=5.7 \lambda^2 \eta G_F^2 T_B^9 \, ,~~~~~~~~~~~~~~~~
\end{eqnarray}
where $\eta$ is a colinearity reduction factor $\eta=\langle p^\mu q_\mu/(E_p E_q)\rangle$,
which is about $.2$ on the axis, at the injection radius, in the sketch of fig. 7, and
is rapidly decreasing as $\theta$ increases. Here the prefactor $4$ comes from the relation of
cross-sections, $\sigma_{\nu+\bar \nu \rightarrow e^+ +e^-}=4 
\sigma_{ e^+ +e^- \rightarrow \nu+\bar \nu}$. Taking $\lambda=.1$, we find the
rate is much less than the energy deposition rate in the scattering reactions (\ref{heat}).

Finally we estimate the rate of radial momentum transfer to the outer layer through
the scattering process as $\approx 3 \times$ the rate of energy transfer. With the
parameters at hand this leads to rough balance with the oppositely directed
rate of momentum deposition due to gravity

In this paper, we have more than once remarked on the role of $\nu$-lepton reactions as
a source of neutrino opacity that can affect the location and temperature
of the $\nu$ surface. Since this issue appears not to have a qualitative effect
on our results, we have not developed it in detail. Here we note that
ref. \cite{nary} and others include only the leptonic reaction included
in the opacity was $\bar \nu_e +\nu_e 
\rightarrow e^+ + e^-$. However, in conditions in which all three favors of $\nu, \bar \nu$
are trapped one needs to take into account all ($\nu$+lepton) reactions, and
for a $\nu$ of any flavor there are 11 separate reactions.  As it turns out, leptonic trapping does not quite occur in the 
present scenario although it could in extreme cases of energy output of the disk.

We have also looked at the same set of processes as a means of establishing
the plasma halo in the first place (or to put it another way, as a means of heating 
a region in which $\rho_{\rm nuc}<10^8$, so that reactions on nucleons are
inconsequential). We find that through the chain of first creating some pairs
from the annihilation process after which the dominant $e-\nu$ heating
takes over, temperatures approaching the steady state temperature are reached
promptly.

\end{document}